\documentclass[12pt,a4paper]{article}%
\pdfoutput=1
\usepackage{cite}
\usepackage{calc}
\usepackage[pdftex]{color,graphicx}
\usepackage{fullpage,epsf, amssymb} 
\usepackage{mathrsfs, epstopdf, amsmath} 
\usepackage{hyperref}
\usepackage{multirow}
\usepackage{braket}
\usepackage{slashed}
\usepackage{appendix}
\usepackage{amsfonts}
\usepackage[pdftex]{color,graphicx}

\graphicspath{{figs/}}

\numberwithin{equation}{section} 
\newcommand{\hhref}[1]{\href{http://arxiv.org/abs/#1}{arXiv:#1}}
\newcommand{\beq}{\begin{equation}}
\newcommand{\eeq}{\end{equation}}

\graphicspath{{figs/}}

\begin{document}
\hfill{CERN-PH-TH/2012-313, DFPD 2012 TH-20, CP3-12-47}

\renewcommand{\thefootnote}{\fnsymbol{footnote}}
\color{black}
\begin{center}
{\Large \bf Lifting degeneracies in Higgs couplings using\\ \vspace{0.25cm} single top production in association with a Higgs boson
}\\[5mm]
\bigskip\color{black}
{{\bf Marco Farina$^{a}$,    Christophe Grojean$^{b,c}$, Fabio Maltoni$^{d}$, \\ \vspace{0.2cm} Ennio Salvioni$^{c,e}$ and Andrea Thamm$^{c,f}$  \footnote{Email: \url{mf627@cornell.edu},~ \url{christophe.grojean@cern.ch},~\url{fabio.maltoni@uclouvain.be}, \\ ~ \url{ennio.salvioni@cern.ch},~ \url{andrea.thamm@cern.ch} }}
} \\[5mm]
{\it  (a)  Department of Physics, LEPP, Cornell University, Ithaca, NY 14853, USA}\\[3mm]
{\it  (b)  ICREA at IFAE, Universitat Aut\`onoma de Barcelona, E-08193 Bellaterra, Spain}\\[3mm]
{\it  (c)  Theory Division, Physics Department, CERN, CH-1211 Geneva 23, Switzerland}\\[3mm]
{\it  (d) Centre for Cosmology, Particle Physics and Phenomenology (CP3), Universit\'e Catholique de Louvain, Chemin du Cyclotron 2, B-1348 Louvain-la-Neuve, Belgium}\\[3mm]
{\it  (e)  Dipartimento di Fisica e Astronomia, Universit\`a di Padova and INFN,\\[1.5mm] Via Marzolo 8, I-35131 Padova, Italy}\\[3mm]
{\it  (f)  Institut de Th\'eorie des Ph\'enom\`enes Physiques, EPFL,  CH-1015 Lausanne, Switzerland}\\[3mm]
\end{center}
\bigskip
\centerline{\bf Abstract}
\begin{quote}
Current Higgs data show an ambiguity in the value of the Yukawa couplings to quarks and leptons. Not so much because of still large uncertainties in the measurements but as the result of several almost degenerate minima in the coupling profile likelihood function.
To break these degeneracies, it is important to identify and measure processes where the Higgs coupling to fermions interferes with other coupling(s). The most prominent example, the decay of $h \to \gamma \gamma$, is not sufficient to give a definitive answer.  In this paper, we argue that $t$-channel single top production in association with a Higgs boson, with $h\to b\bar b$, can provide the necessary information to lift the remaining degeneracy in the top Yukawa. Within the Standard Model, the total rate is highly reduced due to an almost perfect destructive interference in the hard process, $W b \rightarrow t h$. We first show that for non-standard couplings the cross section can be reliably computed without worrying about corrections from physics beyond the cutoff scale $\Lambda\gtrsim 10\,\mathrm{TeV}$, and that it can be enhanced by more than one order of magnitude compared to the SM. We then study the signal $ p p \rightarrow t h j (b)$  with 3 and 4 $b$'s in the final state, and its main backgrounds at the LHC. We find the $8$ TeV run dataset to be sensitive to the sign of the anomalous top Yukawa coupling, while already a moderate integrated luminosity at $14$ TeV should lift the degeneracy completely.
\end{quote}

\renewcommand{\thefootnote}{\arabic{footnote}}\setcounter{footnote}{0}

\enlargethispage{1cm}

\pagestyle{empty}

\newpage

\section{Introduction} \label{sec:intro}
\setcounter{page}{1}
\pagestyle{plain}

After 48 years of desperate searches, the most wanted elementary particle, the Higgs boson, or something that wickedly looks like it, has finally been caught by the ATLAS and CMS experiments~\cite{ATLASdiscovery,CMSdiscovery}. Uncertainties concerning its spin and CP properties remain and will be subject to intense experimental scrutiny in the present and forthcoming run of the LHC. Concurrently, an important program has been launched to measure its couplings to other known elementary particles of the Standard Model (SM). The goal is not so much to determine a few further unknown parameters of the SM but to understand the underlying structures of the laws of physics at high energy: if the SM were to be valid up to the scale of quantum gravity, the couplings of the Higgs boson would be uniquely fixed in terms of other already known and well-measured quantities. On the contrary, any deviation in these couplings, for instance of the order of 20\%, would unambiguously signal new physics at a scale below 5~TeV.

The study of the LHC sensitivity to the Higgs couplings has been initiated in Refs.~\cite{Zeppenfeld:2000td,Duehrssen,Duhrssen:2004cv,Lafaye:2009vr}.
Upon the first and still incomplete measurements reported by both ATLAS and CMS as well as by the Tevatron experiments, a simple methodology inspired by a chiral effective Lagrangian approach has been developed in Refs.~\cite{Carmi:2012yp,Azatov:2012bz,Espinosa:2012ir} in order to quantify to which extent the Higgs boson is really fulfilling the role it has been devoted to in the SM, namely the screening of scattering amplitudes involving massive bosons and fermions at high energy.

At the LHC, the main production channel of the Higgs boson as well as its cleanest decay mode proceed through purely quantum mechanical processes and rely on couplings to massless gluons and photons that are vanishing in the Born approximation. This results in an ambiguity in the value of the tree-level Higgs couplings since the coupling likelihood function exhibits several and almost degenerate minima (see e.g. Refs.~\cite{Lafaye:2009vr,Azatov:2012bz,Espinosa:2012ir}). It has been emphasized~\cite{Espinosa:2012ir, Azatov:2012rd} that these degeneracies are likely to remain even after the on-going analyses will be extended to the whole 8 TeV dataset. Measuring processes involving real top quarks in the final state will bring invaluable information. With the largest rate, the Higgs production in association with a top pair is a golden channel and has received great attention by the experimental~\cite{ATLAStth, CMStth} as well as theoretical~\cite{Beenakker:2001rj,Dawson:2002tg, Frederix:2011zi, Garzelli:2011vp,Degrande:2012gr} communities.

In this paper we argue that, even though subleading, Higgs boson production in association with a single top quark can also bring valuable information, in particular regarding the {\it sign} of the top Yukawa coupling\footnote{The sign of the top Yukawa coupling is not physical by itself, but the relative sign compared to the Higgs coupling to gauge bosons (we take the latter to be positive) is physical.}. This is because an almost totally destructive interference between two large contributions, one where the Higgs couples to a space-like $W$ boson and the other where it couples to the top quark, takes place in the SM. This fact can be exploited to probe deviations in the Higgs coupling structure, which will inevitably jeopardize perturbative unitarity at high energy and lead to a striking enhancement of the cross section compared to the SM.
We discuss how this enhancement can be used to extract information on the sign of the top Yukawa coupling and we show that $th$ production can be used to lift the degeneracy plaguing the Higgs coupling fit of the LHC data. While a moderate integrated luminosity at $14$ TeV should allow us to make a conclusive statement, we point out that already with the full 2012 luminosity, corresponding to $\sim 25\,\mathrm{fb}^{-1}$ per experiment, an interesting sensitivity on the sign of the top Yukawa could be reached.


In our  study we focus on the decay of the Higgs into $b\bar{b}$, updating the early analysis of Ref.~\cite{Maltoni:2001hu} (see also Refs.~\cite{Tait:2000sh,Barger:2009ky}). This choice leads to an experimental signature (lepton + missing energy + multijets, among which $\geq 3$ are $b$-jets) which is very similar to the one ATLAS and CMS have already analyzed in their searches for $t\bar{t}h$ production~\cite{ATLAStth, CMStth}. In this respect we believe that the experimental collaborations could easily perform the analysis we propose here in the very near future, thus adding new important information to the challenge of identifying the true nature of the recently discovered particle.

The large enhancement of the $th$ cross section for nonstandard Higgs couplings is associated to the growth of the scattering amplitude at high energy, which in turn implies that perturbative unitarity is lost at some UV scale $\Lambda$. We estimate $\Lambda$, which acts as the cutoff of our effective theory, to be at least of $\mathcal{O}(10)$ TeV and thus above the energy scales that the LHC will be able to probe. In fact, the $th$ invariant mass distribution in LHC collisions essentially vanishes above 1~TeV, therefore we can safely conclude that our analysis remains insensitive to UV physics above the cutoff scale.


Our paper is structured as follows: we start by introducing the general features of the $th$ process and discussing its implications, including an estimate of the scale where perturbative unitarity is lost, in Section~\ref{sec:main}. We proceed in Section~\ref{sec:signals} to the analysis of the signal and of the main backgrounds at the LHC, performing a parton-level simulation. In Section~\ref{sec:implications}, we discuss the implications on the determination of the Higgs parameters. Finally, we conclude in Section~\ref{sec:conclusions}. Unless otherwise specified, the Higgs mass is assumed to be $m_{h}=125\,\mathrm{GeV}$ throughout this work.  For the top mass we take $m_{t}=173\,\mathrm{GeV}$. Finally, the shorthand $th$ is always understood to include also the charge-conjugated case where $t$ is replaced by $\bar{t}$. Therefore all our cross sections include both $t$ and $\bar t$ production.

\section{Single top and Higgs associated production} \label{sec:main}

\begin{figure}[t]
 \begin{center}
\includegraphics[width=.85\textwidth]{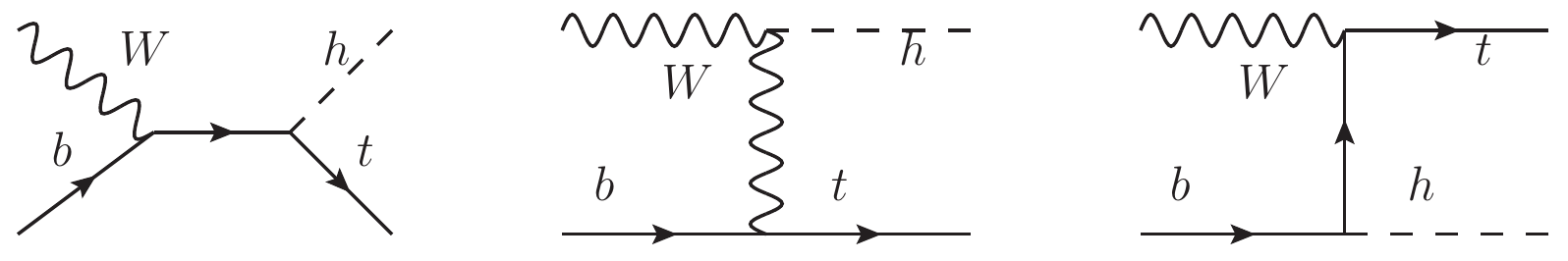}
 \end{center}
 \caption{Feynman diagrams contributing to the partonic process $Wb \rightarrow th$.}
\label{fig:FeynmanDiagrams}
\end{figure}

The Feynman diagrams contributing to the core process $Wb\to th$ are shown in Fig.~\ref{fig:FeynmanDiagrams}. The diagram where the Higgs is emitted from a $b$ leg is suppressed by the bottom Yukawa, and will be consistently neglected in our study. In the $th$ production process at the LHC the initial $W$ is radiated from a quark in the proton, and is thus spacelike. However, at high energy the effective $W$ approximation~\cite{Dawson:1984gx,Kane:1984bb} holds, which allows us to factorize the process into the emission of an approximately on-shell $W$ from the quark times its hard scattering with a bottom. Thus it makes sense to discuss the amplitude for $Wb\to th$ at high energies assuming the initial $W$ to be on-shell, in order to gain an approximate understanding of the full picture.

In the high-energy, hard-scattering regime, where $s, -t, -u \gg m_{t}^{2},m_{W}^{2},m_{h}^{2}$, the amplitude for $W_{L}b\to th$ (the longitudinal polarization dominates at large $s$) reads\footnote{We take final momenta outgoing, and define $s=(p_{W}+p_{b})^{2}$, $t=(p_{W}-p_{h})^{2}$. $\varphi$ is the azimuthal angle around the $z$ axis, which is taken parallel to the direction of motion of the incoming $W$.}
\begin{equation} \label{hard scattering amp}
\mathcal{A}= \frac{g}{\sqrt{2}}\left[(c_{F}-c_{V})\frac{m_{t}\sqrt{s}}{m_{W}v}\,A\left(\frac{t}{s},\varphi; \xi_{t},\xi_{b}\right)+\left(c_{V}\,\frac{2m_{W}}{v}\frac{s}{t}+(2c_{F}-c_{V})\,\frac{m_{t}^{2}}{m_{W}v}\right)\,B\left(\frac{t}{s},\varphi; \xi_{t},\xi_{b}\right)\right]\,,
\end{equation}
where we have omitted terms that vanish in the high-energy limit and, for simplicity, also neglected the Higgs mass in addition to setting $m_{b}=0$. The generalized couplings of the Higgs are defined as $c_{V}\equiv g_{hWW}/g_{hWW}^{SM}$ and $c_{F}\equiv g_{ht\bar{t}}/g_{ht\bar{t}}^{SM}$. The functions $A,B$ are given by
\begin{align} \label{A}
 A\left(t/s,\varphi;\xi_{t},\xi_{b}\right) =\,\,& \xi_{t}^{\dagger}\begin{pmatrix}
 -t/s & 0 \\
 - e^{i\varphi}\sqrt{-\frac{t}{s}\left(1+\frac{t}{s}\right)} & 0 \end{pmatrix} \xi_{b}\quad \longrightarrow \quad  \begin{pmatrix}
 0 & \,0 \\
 - e^{i\varphi}\sqrt{-t/s} & \,0 \end{pmatrix}\,, \\ \label{B}
B\left(t/s,\varphi;\xi_{t},\xi_{b}\right) =\,\,& \xi_{t}^{\dagger}\begin{pmatrix}
 1+t/s & \,\,\,\; 0 \\
 e^{i\varphi}\sqrt{-\frac{t}{s}\left(1+\frac{t}{s}\right)} & \,\,\,\; 0 \end{pmatrix} \xi_{b} \quad \longrightarrow \quad  \begin{pmatrix}
 e^{i\varphi}\sqrt{1+t/s} & 0 \\
 0 & 0 \end{pmatrix}\,,
\end{align}
where in the rightmost term of each line we have chosen a specific basis for the spinors, namely
\begin{equation} \label{spinorbasis}
\xi_{b}^{L} = \begin{pmatrix} 1 \\ 0 \end{pmatrix},\quad \xi_{b}^{R} = \begin{pmatrix} 0 \\ 1 \end{pmatrix}; \qquad \xi_{t}^{L}= \begin{pmatrix} e^{-i\varphi}\sqrt{1+t/s} \\
\sqrt{-t/s} \end{pmatrix}\,,\quad \xi_{t}^{R}= \begin{pmatrix} - e^{-i\varphi}\sqrt{-t/s} \\
\sqrt{1+t/s} \end{pmatrix}\,,
\end{equation}
which correspond to the chiral states $\{F_{L},F_{R}\}$ ($F=b,t$) in the $m_{F}\to 0$ limit\footnote{However, note that the limit $m_{t}\to 0$ does not interest us here.}. The amplitudes involving the helicity state $\xi_{b}^{R}$, which is identified with a right-handed bottom since we are assuming $m_{b}=0$, exactly vanish due to the $V-A$ structure of the couplings of the $W$ to fermions. From Eq.~\eqref{hard scattering amp} we see that when $c_{V}\neq c_{F}$ the amplitude grows with energy like $\sqrt{s}$ and is enhanced compared to the case $c_{V} = c_{F}$ (which includes the SM), where the amplitude is constant in the large $s$ limit. The non-cancellation of the terms in the amplitude growing with energy is at the origin of the striking enhancement of the cross section when $c_{V}\neq c_{F}$.

\begin{figure}[t]
 \begin{center}
\includegraphics[width=.55\textwidth]{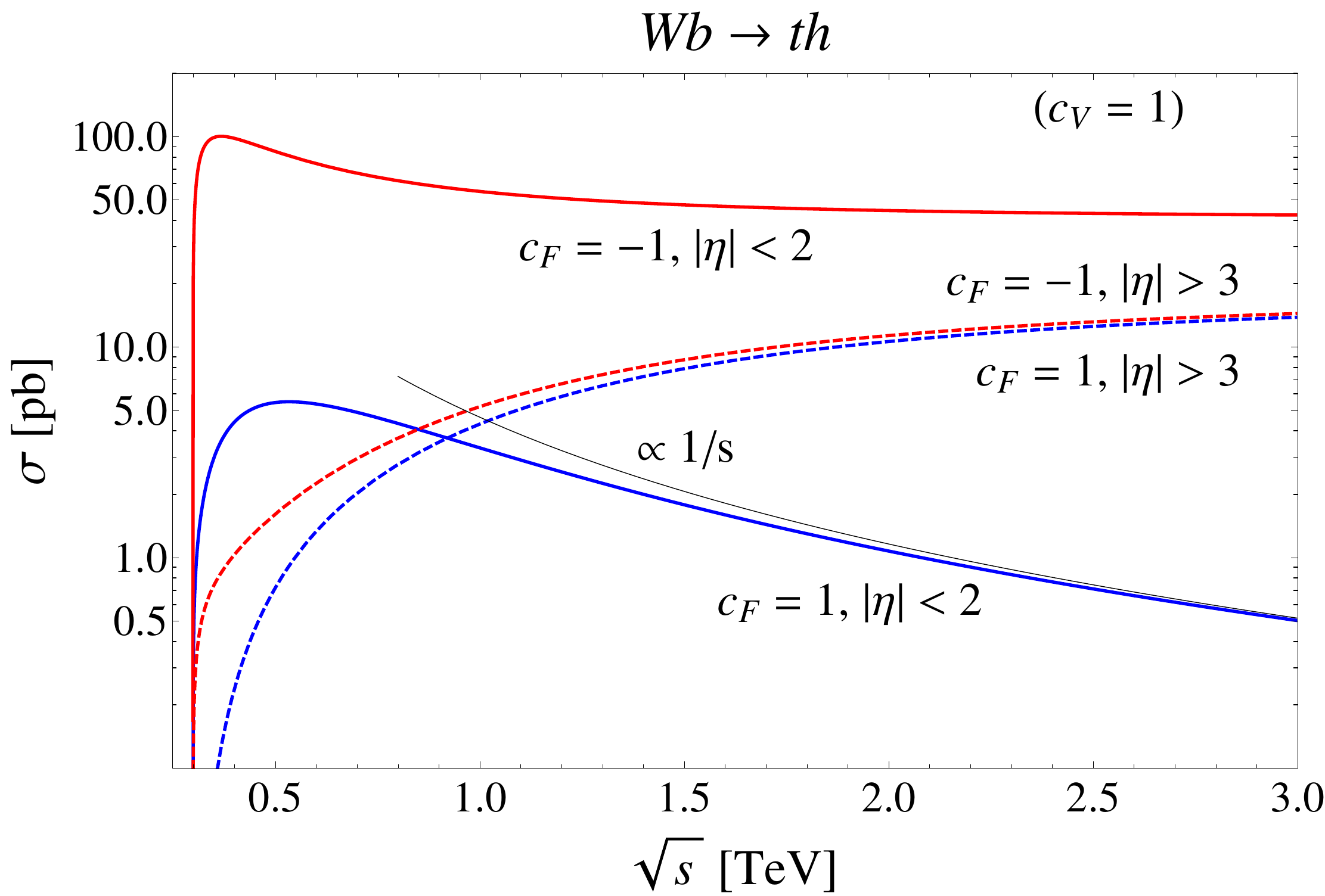}
 \end{center}
 \caption{Partonic cross sections for the process $Wb\to th$ as a function of the center of mass energy $\sqrt{s}$. The parameter $c_{V}$ is set to $1$. The hard scattering cross section is defined by a cut $|\eta|<2$: the large enhancement obtained for $c_{F} = - c_{V}$ with respect to the SM case is evident. The forward cross section, defined by a cut $|\eta|>3$, is also shown (dashed curves).}
\label{fig:partonic cs}
\end{figure}

The cross section for $Wb\to th$ is shown as a function of the center of mass energy in Fig.~\ref{fig:partonic cs}. The large enhancement of the hard scattering cross section (defined by a centrality cut $|\eta|<2$) for $c_{F} = - c_{V}$ is evident.\footnote{Incidentally, we note that the cross section shows another feature, a Coulomb enhancement at small $|t|$ due to the diagram with a $W$ exchange in the $t$-channel. As can be read off Fig.~\ref{fig:partonic cs} the forward cross section tends to a constant limit for large $s$, which can be computed in a simple way in terms of the parameter $c_{V}$ alone and is insensitive to the value of $c_{F}$. A short discussion of the forward cross section is contained in Appendix~\ref{App: fw scatt}.} At large energies, the amplitude is constant for $c_{V} = c_{F}$ and thus the cross section vanishes as $\sim 1/s$. On the other hand, when $c_{F}\neq c_{V}$ the amplitude grows with energy like $\sqrt{s}$ and as a consequence the cross section tends to a constant for large $s$. It is easy to compute this asymptotic value of the cross section: squaring the leading term of the amplitude in Eq.~\eqref{hard scattering amp}, summing and averaging over polarizations and integrating over $t$ we find
\begin{equation} \label{hard cs}
\sigma(|\eta|<\tilde{\eta},s\to \infty) \simeq \frac{g^{2}(c_{F}-c_{V})^{2}m_{t}^{2}}{384\pi\, m_{W}^{2}v^{2}}\tanh \tilde{\eta}\,.
\end{equation}
This simple formula gives accurate results: for example for $\sqrt{s}= 5 \, \mathrm{TeV}$, \mbox{$c_{V}= -c_{F} = 1$} and a centrality cut\footnote{Note that for the expression in Eq.~\eqref{hard cs} to be reliable, $\tilde{\eta}$ cannot be too large. In fact, as already mentioned, in the forward region the cross section has a Coulomb enhancement which is not captured by the approximations we made here. See also Appendix~\ref{App: fw scatt}.} $|\eta|<2$ we find that the cross section computed without any approximations is $\sigma_{\mathrm{full}}(|\eta|< 2) = 41.3\,\mathrm{pb}$, whereas \mbox{$\sigma(|\eta|< 2,s\to \infty) = 40.7\,\mathrm{pb}\,$}.

Since for $c_{V}\neq c_{F}$ the hard scattering amplitude grows with energy, perturbative unitarity will be lost at some cutoff scale $\Lambda$, which we now estimate. In the spinor basis of Eq.~\eqref{spinorbasis}, only one $s$-wave amplitude is non-vanishing
\begin{equation} \label{hardscatt-leadingpart}
a_{0} = \frac{1}{16\pi\sqrt{2} \sqrt{s}}(c_{F}-c_{V})\frac{g m_{t}}{m_{W}v} \int_{-s}^{0}A(t/s, \varphi; \xi_{t}^{R},\xi_{b}^{L}) = -\frac{1}{24\sqrt{2}\pi}(c_{F}-c_{V})\frac{g m_{t}\sqrt{s}}{m_{W}v}e^{i\varphi}
\end{equation}
from which, imposing the condition $|a_{0}|<1$, we find that perturbative unitarity is violated at a scale $\sqrt{s}\simeq \Lambda$ with
\begin{equation}\label{cutoff}
\Lambda = 12\sqrt{2}\pi \frac{v^{2}}{m_{t}\left|c_{F}-c_{V}\right|}\,.
\end{equation}
For example, for $c_{V}= -c_{F} = 1$ the cutoff is $\Lambda\simeq 9.3\,\mathrm{TeV}$. One may worry about other processes involving top quarks, in which perturbative unitarity could be lost at a scale lower than the one in Eq.~\eqref{cutoff} for $c_{F}<0$. A relevant and often mentioned process is $W_{L}^{+}W_{L}^{-}\to t\bar{t}$, for which we find
$
\Lambda = 16\pi v^{2}/(m_{t}\left|1-c_{V}c_{F}\right|)\,.
$
For $c_{V}=-c_{F}=1$ this formula yields $8.8\,\mathrm{TeV}$, essentially the same cutoff scale we found for $W_{L}b\to th$. For previous discussions of perturbative unitarity breakdown in processes with external fermions, see Refs.~\cite{Appelquist:1987cf, Maltoni:2001dc}.

\begin{figure}[t]
 \begin{center}
\includegraphics[width=.55\textwidth]{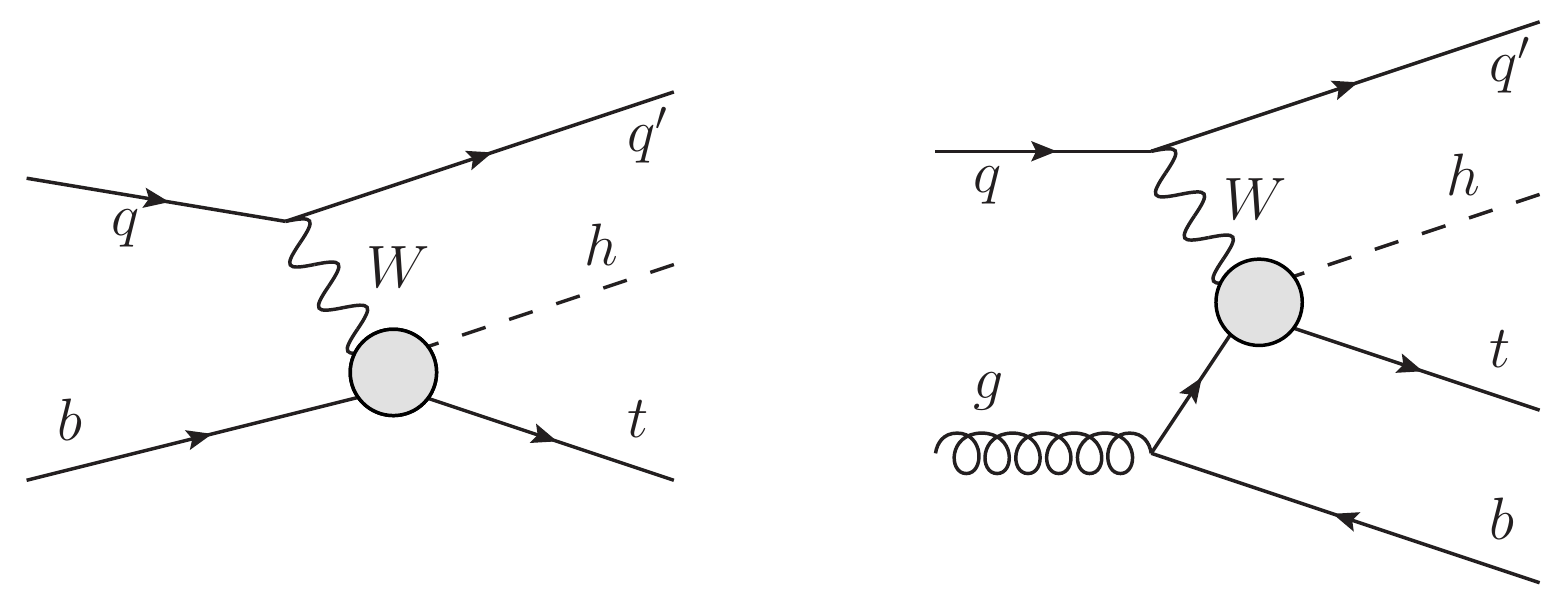}
 \end{center}
 \caption{Feynman diagrams for the processes $p p \rightarrow thj$ and $p p \rightarrow thjb$.}
\label{fig:feynman hadronic}
\end{figure}

Having analyzed the behavior of the partonic cross section, we can now turn our attention to single top and Higgs associated production in hadron collisions. At the LHC, $t$-channel single top production goes through an initial-state gluon splitting into a $b \overline{b}$ pair. Such a process can be efficiently described by a $5$-flavor scheme where $b$'s are in the initial state and described by a perturbative $b$ PDF, Fig.~\ref{fig:feynman hadronic}(a). In this scheme, the non-collinearly enhanced contribution, where the spectator $b$ (i.e. the one not struck by the $W$ boson) is central and at high $p_{T}$ (see Fig.~\ref{fig:feynman hadronic}(b)), is moved to the next-to-leading order term. This contribution, which we indicate with $pp\to thjb$, is finite and can be easily calculated at tree-level, contributing to a final state signature with an extra $b$-jet, a useful handle to suppress the background. In Table~\ref{tab:inclusive cs} we present the rates for $th$ production in the $5$-flavor scheme, fully inclusive as well as with the requirement of the extra $b$ to be in the tagging region, for $8$ and $14$ TeV, in the $c_{V} = 1, c_{F} \pm 1$ cases. Our analysis in Section~\ref{sec:signals} will consider both processes, which lead to final states containing $3$ and $4$ $b$-jets respectively, once the decay of the Higgs to $b\bar{b}$ is taken into account. The cross sections in Table~\ref{tab:inclusive cs} were computed using {\sc MadGraph~5}\cite{Alwall:2011uj} with CTEQ6L1 PDFs~\cite{Pumplin:2002vw}, setting the factorization and renormalization scales to the default event-by-event {\sc MadGraph~5} value. As an estimate of the theoretical uncertainty on the signal, we have computed the fully inclusive cross sections at NLO in QCD, in the $5$-flavor scheme, using {\sc aMC@NLO}~\cite{Frederix:2009yq,Hirschi:2011pa,amcatnlo} and CTEQ6M PDFs~\cite{Pumplin:2002vw}. The results are reported in Table~\ref{tab:NLO cs}, where the uncertainties correspond to variations of the factorization and renormalization scales with $\mu_F=\mu_R$ around $\mu = (m_{t}+m_{h})/2$ from $\mu/2$ to $2\mu$.  The NLO cross sections appear to be extremely stable under radiative corrections and therefore we deem the theory uncertainty of the signal rates in our analysis negligible.
\begin{table}[t]
\begin{center}
{\small
\begin{tabular}{c|cc|cc}
\hline \hline
 &&&& \\[-0.3cm]
 & \multicolumn{2}{c}{$\sigma^{{\rm LO}}(pp\rightarrow thj) \, \, [\mathrm{fb}]$} &\multicolumn{2}{|c}{$\sigma^{{\rm LO}}(pp\rightarrow thjb) \, \, [\mathrm{fb}]$}\\[0.04cm]
 & $c_F=1$		& $c_F=-1 $ & $c_F=1$		& $c_F=-1 $\\[0.02cm]
 \hline
 &&&& \\[-0.3cm]
 $8 \,$ TeV & $17.4$		& $252.7$ & $5.4$		& $79.2$\\[0.02cm]
 $14 \,$ TeV & $80.4$		& $1042$ &  $26.9$		& $363.5$\\[0.02cm]
 \hline \hline
\end{tabular}
} \\[0.1cm]
\caption{Leading-order cross sections for the processes $pp\rightarrow thj$ and $pp\rightarrow thjb$ (with $p_T^b>25$ GeV and $|\eta^{b}|<2.5$) at the LHC. The parameter $c_{V}$ has been set to $1$.}
\label{tab:inclusive cs}
\end{center}
\end{table}
%
%
%
\begin{table}[t]
\begin{center}
{\small
\begin{tabular}{c|cc}
\hline \hline
 && \\[-0.3cm]
 & \multicolumn{2}{c}{$\sigma^{{\rm NLO}}(pp\rightarrow thj) \, \, [\mathrm{fb}]$}\\[0.04cm]
   & $c_F=1$ & $c_F=-1 $ \\[2pt]
 \hline
 && \\[-6pt]
 $8 \,$ TeV & $18.28^{+0.42}_{-0.38} $		& $233.8^{+4.6}_{-0.}$ \\[10pt]
 $14 \,$ TeV & $88.2^{+1.7}_{-0.}$ & $982^{+28}_{-0}$ \\[4pt]
\hline \hline
\end{tabular}
} \\[0.1cm]
\caption{Cross sections at NLO in QCD for the process $pp\rightarrow thj$ at the LHC. The parameter $c_{V}$ has been set to $1$.}
\label{tab:NLO cs}
\end{center}
\end{table}

\begin{figure}[t]
 \begin{center}
\includegraphics[width=.5\textwidth]{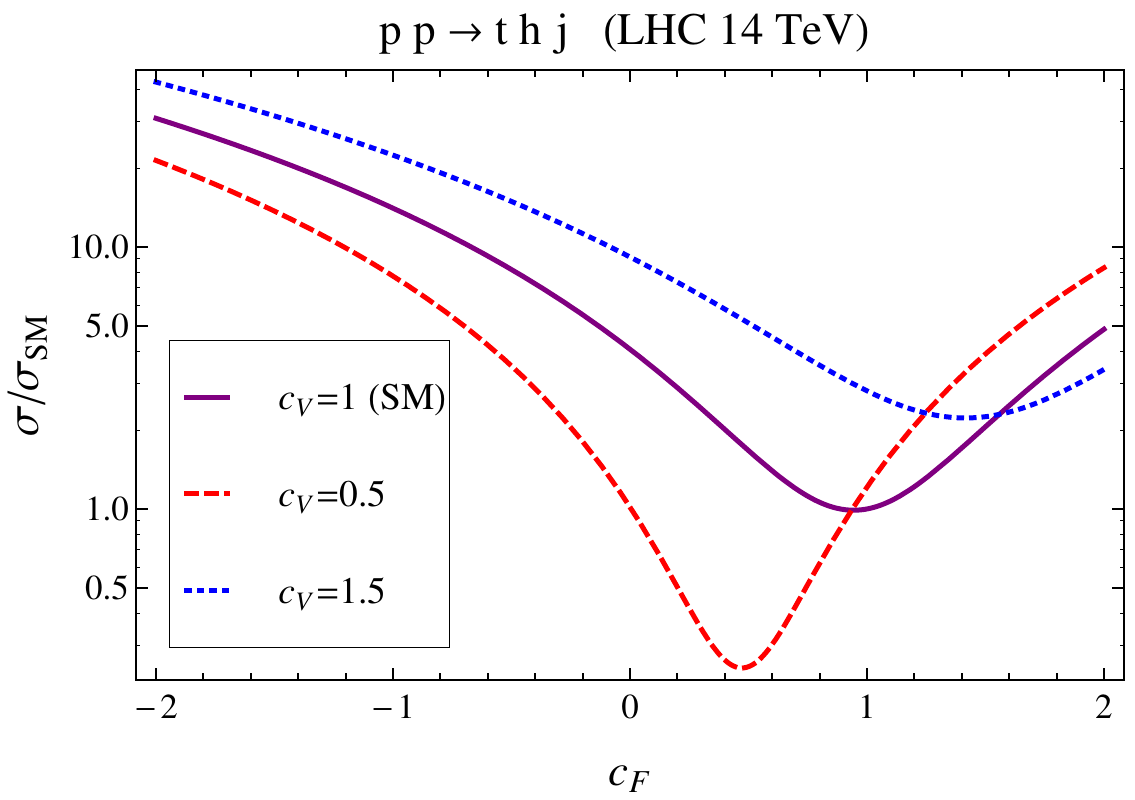}
 \end{center}
 \caption{Cross section for $pp\rightarrow thj$ at $14$ TeV normalized to the SM one, as a function of $c_F$ for three choices of $c_{V}$. Solid, dashed and dotted lines correspond to $c_V=1, \, 0.5$ and $1.5$ respectively.}
\label{fig:sigmaRatio}
\end{figure}

The striking enhancement of the hadronic cross section for $c_{F}\neq c_{V}$ is shown in Fig.~\ref{fig:sigmaRatio}, where $\sigma(p p \rightarrow t h j)$ for an LHC energy of $14\,\mathrm{TeV}$, normalized to its SM value, is displayed as a function of $c_{F}$ for three different choices of $c_{V}$ (very similar plots are obtained considering $8\,\mathrm{TeV}$ and/or the $pp\to thjb$ process). For example, for a standard $hWW$ coupling, i.e. $c_{V}=1$, a top Yukawa with equal magnitude and opposite sign with respect to the standard one ($c_{F}=-1$) yields an enhancement of the cross section of more than a factor 10.

\begin{figure}[t]
 \begin{center}
\includegraphics[width=.45\textwidth]{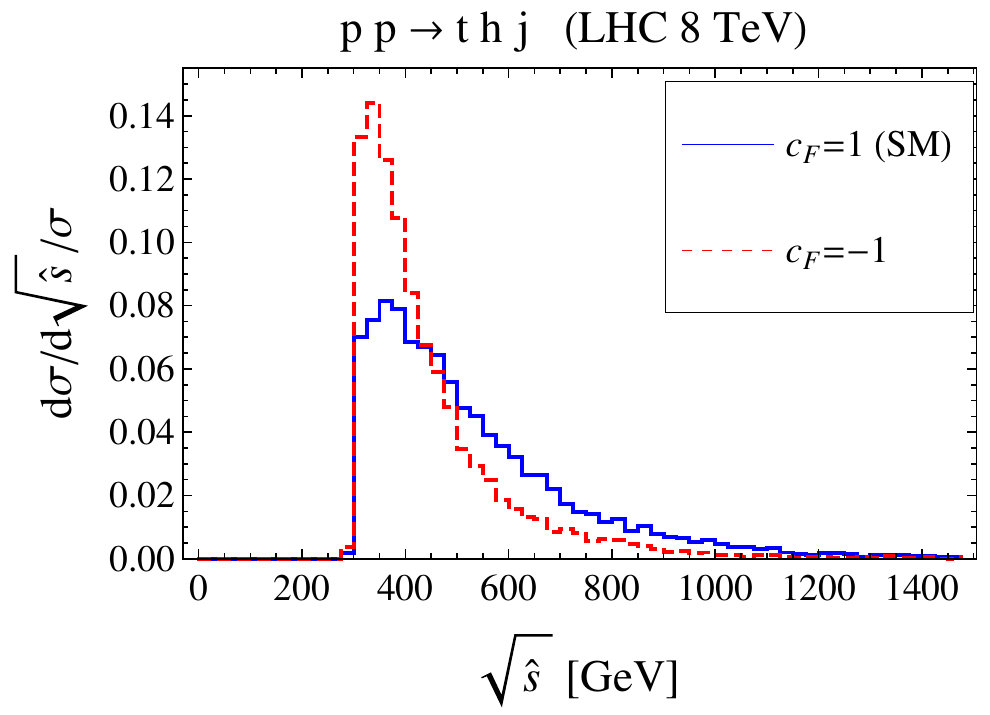}\hspace{1cm}\includegraphics[width=.45\textwidth]{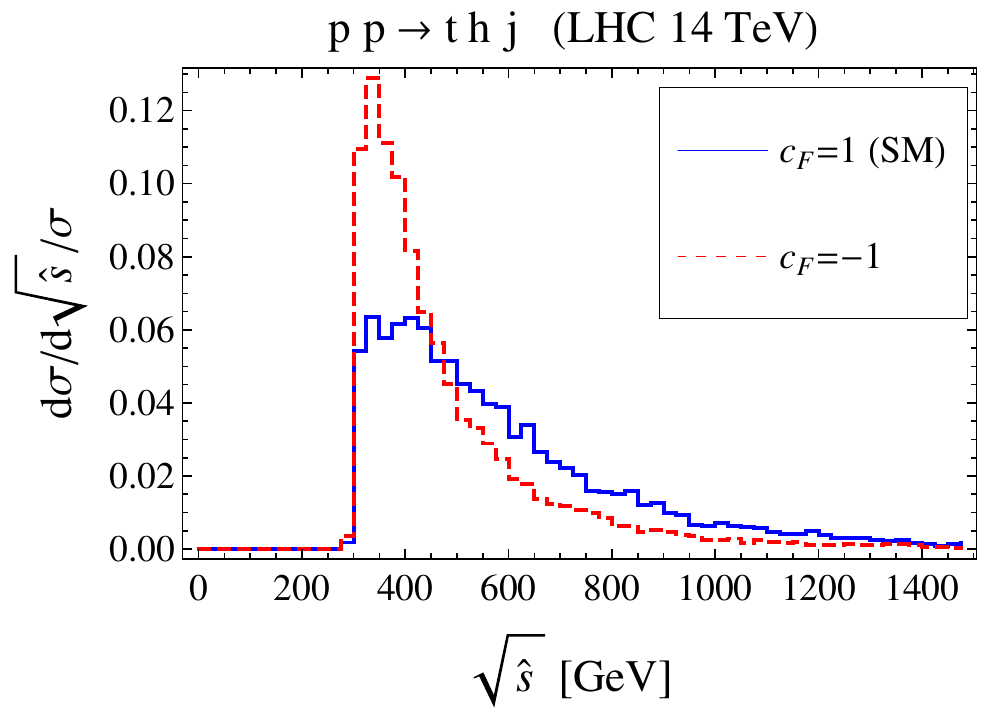}
 \vspace{-.5cm}
 \end{center}
 \caption{Histograms of normalized $p p \rightarrow thj$ cross section as a function of the center of mass energy of the hard scattering process $Wb\to th$. The left panel is for $8$ TeV, the right one for $14$ TeV.}
\label{fig:histoS}
\end{figure}

As noted above, perturbative unitarity in $Wb\to th$ scattering is lost at a scale $\Lambda\gtrsim 10\, \mathrm{TeV}$ for $c_{V},c_{F}\sim \mathcal{O}(1)$. Figure~\ref{fig:histoS} clearly shows that after convolution with the PDFs the contribution of the region $\sqrt{\hat{s}} \gtrsim 1\,\mathrm{TeV}$, where $\sqrt{\hat{s}}$ is the center of mass energy of the $th$ system, to the hadronic cross section is negligible. This implies that our perturbative computations can be fully trusted. Indeed Fig.~\ref{fig:histoS} demonstrates that the relative contribution to the cross section from large values of $\sqrt{\hat{s}}$ is more sizable in the SM than for $c_{F}\neq c_{V}$. This is compatible with the different behaviors of the partonic cross section in the two cases, shown in Fig.~\ref{fig:partonic cs}.


\section{Signal and background study}\label{sec:signals}

\subsection{Parton-level simulation} \label{subsec:simulation}

Signal and background events have been generated at the parton level using {\sc MadGraph~5} with CTEQ6L1 PDFs, setting the factorization and renormalization scales to the default event-by-event {\sc MadGraph~5} value. Jets are defined at the parton level. In order to take showering, hadronization, detector and reconstruction effects minimally into account, we smear the $p_{T}$ of the jets uniformly in $\eta$ using a jet energy resolution defined by
\begin{equation} \label{JetEnRes}
\frac{\sigma (p_T)}{p_T}=\frac{a}{p_T} \oplus \frac{b}{\sqrt{p_T}}\oplus c \, ,
\end{equation}
where the parameters are taken to be $a = 2$, $b = 0.7$ and $c = 0.06$. With these choices, Eq.~\eqref{JetEnRes} is compatible with the results of the ATLAS jet energy resolution study of Ref.~\cite{Aad:2012ag} (see Fig.~9 there). The jet 4-momentum is then rescaled by a factor $p_{T}^{smeared}/p_{T}\,$.
%
%
The acceptance cuts reported in Table~\ref{tab:cuts}, chosen following the ATLAS $t\bar{t}h$ analysis~\cite{ATLAStth}, are applied on the physical objects. We do not require any acceptance cut on the missing transverse energy.
\begin{table}[h]
\begin{center}
{\small
\begin{tabular}{c|cccccc}
\hline  \hline
 &&&&&& \\[-0.3cm]
Cut		& $p_{T}^{b} > $  & $p_{T}^{\ell} > $ & $p_{T}^{j} > $ & $|\eta^{b, \ell}| < $  &  $|\eta^{j}| < $  &  $\Delta R_{ij} > $  \\[0.02cm]
 \hline
 &&&&&& \\[-0.3cm]
Value 	& $25$ GeV & $25$ GeV & $30$ GeV &  $2.5$ & $5$ & $0.4$   \\[0.02cm]
 \hline  \hline
\end{tabular}
} \\[0.1cm]
\caption{Acceptance cuts applied to the signal and backgrounds at the reconstructed level. The $\Delta R$ requirement applies to all objects.}
\label{tab:cuts}
\end{center}
\end{table}

An object is considered to be missed if it does not pass one of the acceptance cuts. If, in particular, two jets are collinear with $\Delta R<0.4$ we merge them by summing their 4-momenta and we consider them as a single jet when applying further cuts.\footnote{The exception to this procedure is the case where the $b$ coming from a semileptonic top decay is collinear to another jet. Since we are assuming ideal semileptonic top reconstruction (see below), we simply reject the event in this case.} Additionally we require the lepton to be isolated from any jet in the event, including those that do not pass acceptance cuts and therefore are missed.

In all the signal and background processes we consider in this paper, a semileptonically decaying top is present. We assume a $100\%$ efficiency for the reconstruction of this top, which implies an unambiguous identification of the $b$ originating from its decay. This assumption is of course idealized, however the use of a more realistic semileptonic top reconstruction efficiency will only affect the overall normalization of both signal and background, and not their relative values.

Concerning $b$-tagging, we assume the following performance: efficiency $\epsilon_{b}=0.7$, charm mistag probability $\epsilon_{c}=0.2$ and light jet mistag probability $\epsilon_{j} \approx 0.008$~\cite{ATLAStth}. Finally we assume a lepton reconstruction efficiency $\epsilon_{\ell} = 0.9$.
%
%
%

%
%

\subsection{Final state with $3$ $b$-tags}

We start by discussing the 3 $b$-jet final state, which arises from $pp\to thj$ after selecting the Higgs decay into $b\bar{b}$. Requiring the top to decay semileptonically ($t \rightarrow b \ell^{+} \nu$) gives the signature
\begin{equation}
3\,b  + 1 \, \text{forward jet}+\ell^{\pm} + E_T^{miss}.
\end{equation}
We can now turn our attention to the most relevant backgrounds:\footnote{For the sake of readability we do not write the top decay $t \rightarrow b l^{+} \nu$ explicitly, as it is the same for all processes.}
\begin{itemize}
\item $t Z j$, $ Z \rightarrow b \overline{b}$: an irreducible background where a $Z$ boson mimics the Higgs in decaying to $b\overline{b}$.
\item $t b \overline{b} j$: an irreducible QCD background.
\item $t \overline{t}$, $\overline{t} \rightarrow \overline{b} \overline{c} s$: a reducible background where either the $c$ or $s$ are mis-tagged.
\item $t \overline{t} j$,  $\overline{t} \rightarrow \overline{b} \overline{c} s$: also in this case, either the $c$ or $s$ are mis-tagged while the other is missed.
\end{itemize}
As can be seen in Table~\ref{tab:3b backgrounds8} for 8 TeV and in Table~\ref{tab:3b backgrounds14} for 14 TeV, after acceptance cuts and efficiencies the last two backgrounds are extremely large. In particular, their values are larger than those quoted in Ref.~\cite{Maltoni:2001hu}, mainly due to a larger charm mistag rate considered here (we use $\epsilon_{c}=0.2$, whereas Ref.~\cite{Maltoni:2001hu} adopted $\epsilon_{c}=0.1$) and to the fact that we increased the $p_{T}$ threshold for jets, which results in a larger probability of missing a jet from $t\bar{t}j$. The dominance of backgrounds where a $c$ is mistagged suggests that it may be sensible to prefer a $b$-tagging performance with smaller efficiency but higher rejection against charm. However, for definiteness we stick to the numbers reported in Section~\ref{subsec:simulation}, taken from Ref.~\cite{ATLAStth}.

After acceptance cuts and efficiencies, the signal is overwhelmed by the $t\bar{t}$ background not only for the standard case $c_{F}=1$, but even considering the enhanced case $c_{F}=-1$ (we set $c_{V}=1$). Thus, we require a set of additional cuts in order to isolate the signal. These cuts are listed in Tables~\ref{tab:3b backgrounds8}-\ref{tab:3b backgrounds14}, together with the cross sections obtained after their application. The value of each cut is chosen by optimizing the Poisson exclusion limit in the $c_{F}=-1$ case. We remark that since we are assuming ideal top reconstruction, the $b$ coming from the semileptonic top is always assumed to be unambiguously identified, therefore no cut on it is applied beyond the detector ones, neither for the signal nor for the backgrounds.

The first cut we apply requires the $b\overline{b}$ pair to have an invariant mass around $m_h$, which of course helps to eliminate the $tZj$ background. The second cut selects large values for the $bbj$ invariant mass and is effective against the reducible backgrounds, in particular it suppresses enormously $t\bar{t}$, where the jet and 2 $b$'s are decay products of a top and therefore we expect their invariant mass to be close to $m_t$. The last cut singles out a forward jet, which is a distinctive feature of the signal. However, after all cuts the background cross section, completely dominated by $t\bar{t}j$, is still one order of magnitude larger than the signal for $c_{F}=-1$.

\begin{table}[t]
\begin{center}
{\small
\begin{tabular}{c|cc|c|cccc}
\hline \hline
\vspace{-0.35cm}\\
		 & \multicolumn{2}{c|}{Signal} &\multicolumn{5}{c}{Backgrounds} \\[0.02cm]
		
 \hline
  &&&&&& \\[-0.3cm]
	Cuts	& $c_F=1$ & $c_F= -1$  & Total & $t Z j$ & $ t b \overline{b} j$ & $t \overline{t} $  &    $t \overline{t} j$\\[0.02cm]
		
 \hline
 &&&&& \\[-0.3cm]
 Acceptance Cuts + $\epsilon$           				& $0.18$ & $2.88$ &$600.81$ &$0.61$ & $1.01$ & $456.40$ & $142.80$ \\[0.02cm]

$|m_{bb} - m_{h}| < 15$ GeV 		& $0.15$ & $2.55$ &$245.95$ &$0.02$ & $0.11$ & $184.2$ & $61.65$ \\[0.02cm]

$m_{bbj} > 270$ GeV 		                    	& $0.10$ & $2.02$ &$31.78$ &$0.01$ & $0.08$ & $0.$ & $30.68$\\[0.02cm]

 $|\eta^j|>1.7$ 		& $0.08$ & $1.70$ &$17.98$ &$0.01$ & $0.06$ & $0.$ & $17.24$ \\[0.02cm]

\hline
 &&&&& \\[-0.3cm]
Events at $25 \, \mathrm{fb}^{-1}$	                    & $1.9$ & $42.5$ &$449.4$ &$ $ & $$ & $$ & $ $ \\[0.02cm]
\hline \hline
\end{tabular}
} \\[0.1cm]
\caption{Cross sections in~fb for the $3$ $b$-tag case at $8$ TeV. In the event line backgrounds are summed.}
\label{tab:3b backgrounds8}
\end{center}
\end{table}

\begin{table}[t]
\begin{center}
{\small
\begin{tabular}{c|cc|c|cccc}
\hline \hline
\vspace{-0.35cm}\\
		 & \multicolumn{2}{c|}{Signal} &\multicolumn{5}{c}{Backgrounds} \\[0.02cm]
		
 \hline
  &&&&&& \\[-0.3cm]
	Cuts	& $c_F=1$ & $c_F= -1$ & Total & $t Z j$ & $ t b \overline{b} j$ & $t \overline{t} $  &    $t \overline{t} j$  \\[0.02cm]
		
 \hline
 &&&&& \\[-0.3cm]
 Acceptance Cuts + $\epsilon$          				& $0.71$ & $11.55$ &$2448.18$ &$2.29$ & $3.72$ & $1773.35$ & $668.83$ \\[0.02cm]

$|m_{bb} - m_{h}| < 15$ GeV 		& $0.63$ & $10.23$ &$1020.4$ &$0.07$ & $0.38$ & $737.23$ & $282.76$ \\[0.02cm]

 $m_{bbj} > 280$ GeV 		                    	& $0.46$ & $8.59$ &$153.10$ &$0.05$ & $0.31$ & $0.$ & $152.74$ \\[0.02cm]

$|\eta^j|>2$ 		& $0.34$ & $7.12$ &$79.26$ &$0.03$ & $0.24$ & $0.$ & $79.00$ \\[0.02cm]

\hline
 &&&&& \\[-0.3cm]
Events at $25 \, \mathrm{fb}^{-1}$	                    & $8.4$ & $178.0$  &$1981.5$ &$ $ & $$ & $$ & $ $\\[0.02cm]
\hline \hline
\end{tabular}
} \\[0.1cm]
\caption{Cross sections in~fb for the $3$ $b$-tag case at $14$ TeV. In the event line backgrounds are summed.}
\label{tab:3b backgrounds14}
\end{center}
\end{table}

In the last line of Tables~\ref{tab:3b backgrounds8} and~\ref{tab:3b backgrounds14}, we present the number of signal and total background events expected after $25 \,\mathrm{fb}^{-1}$ of integrated luminosity. At 8 TeV, the Poisson exclusion is at $97.4\%$ CL or $2.2\,\sigma$ (by abuse of notation, we are expressing the probability in terms of number of $\sigma$'s, e.g. $2\,\sigma$ approximately corresponds to the $95\%$ CL), while at $14$ TeV it reaches $\sim 4 \, \sigma$.

\subsection{Final state with $4$ $b$-tags}
As suggested in Ref.~\cite{Maltoni:2001hu}, a way to enhance the sensitivity on the $th$ signal is to require an extra $b$, coming from the splitting of an initial gluon: the process of interest is thus $pp\to thjb$. Requiring a semileptonic top and the decay $h\to b\bar{b}$ leads to the signature
\begin{equation}
4\,b + 1 \, \text{forward jet}+\ell^{\pm} + E_T^{miss}.
\end{equation}
Here the main backgrounds are:
\begin{itemize}

\item $t Z \overline{b} j$, $ Z \rightarrow b \overline{b}$: an irreducible background where the $Z$ mimics the Higgs.
\item $t \overline{b} b \overline{b} j$: similarly to the $3$ $b$ case, an irreducible QCD background.
\item $t \overline{t} b \overline{b}$, $\overline{t} \rightarrow \overline{b} j j$: a reducible background where one of the two jets, originating from a hadronically decaying $W$, is missed.
\item $t \overline{t} b \overline{b}$, $\overline{t} \rightarrow \overline{b} \overline{c} s$ (one mistag): here the $c$ or the $s$ is mis-tagged, while either the other one is missed (and one $b$ is not tagged) or one of the $b$'s is missed.
\item $t \overline{t} j$,  $\overline{t} \rightarrow \overline{b} \overline{c} s$ (two mistags): in this case both $c$ and $s$ are mistagged.
\end{itemize}

Looking at Tables~\ref{tab:4b backgrounds8} and~\ref{tab:4b backgrounds14}, we see that requiring 4 $b$-jets allows us to obtain a much larger signal to background ratio after acceptance cuts compared to the $3$ $b$ case. On the other hand, the overall rates are obviously smaller. Analogously to what was done in the $3$ $b$ case, a set of  additional cuts are imposed to enhance the signal. The cuts are listed in Tables~\ref{tab:4b backgrounds8}-\ref{tab:4b backgrounds14}, together with the cross sections obtained after their application. The value of each cut is again chosen by optimizing the Poisson exclusion limit in the $c_{F}=-1$ case.

The first cut requires the invariant mass of one of the 3 $bb$ pairs (we recall that ideal reconstruction of the semileptonic top is assumed) to be inside a window around $m_h$. This helps again to eliminate the $tZ \bar{b}j$ background. The second cut demands all $bb$ invariant masses to be higher than about $100\,\mathrm{GeV}$, and is most effective on $t\bar{t}j$, where the mis-tagged $c$ and $s$, coming from a $W$ decay, have an invariant mass around $m_W$. The last cut requires all 3 $bj$ pairs to have a large invariant mass. This efficiently suppresses the $t\bar{t}b\bar{b}$ backgrounds, for which in most cases at least one $bj$ pair comes from a top decay and thus has an invariant mass $m_{bj}\lesssim \sqrt{m_{t}^{2}-m_{W}^{2}}\sim 150\,\mathrm{GeV}$.

\begin{table}[t]
\begin{center}
{\small
\begin{tabular}{c|cc|c|ccccc}
\hline \hline
\vspace{-0.35cm}\\
		 & \multicolumn{2}{c|}{Signal} &\multicolumn{5}{c}{Backgrounds} \\[0.02cm]
		
 \hline
  &&&&&&& \\[-0.3cm]
	Cuts	& $c_F=1$ & $c_F= -1$ & Total & $t Z \bar{b} j$ & $ t b \bar{b} \bar{b} j$ & $t \bar{t} b \bar{b}$  &  $t \bar{t} b \bar{b}$ (\emph{mis}) &  $t \bar{t} j$  \\[0.02cm]
		
 \hline
 &&&&&& \\[-0.3cm]
  Acceptance Cuts + $\epsilon$             				& $0.043$ & $0.63$ &$7.81$ &$0.11$ & $0.26 $ & $2.66\,(0.48)$ & $2.25$ & $2.54$\\[0.02cm]

$|m_{bb} - m_{h}| < 15$ GeV 		& $0.039$ & $0.58$ &$4.06$  &$0.03$ & $0.08 $ & $0.94\,(0.40)$ & $1.29$ & $1.71$\\[0.02cm]

min $m_{b b} > 110$ GeV 		& $0.023$ & $0.30$ &$0.67$  &$0.002$ & $0.015 $ & $0.20\,(0.18)$ & $0.44$ & $0. $\\[0.02cm]

min $m_{bj} > 180$ GeV 		                    	& $0.008$ & $0.15$ &$0.014$ &$0. $ & $0.007 $ & $0.002\,(0.001)$ & $0.004$ & $0.$ \\[0.02cm]
\hline
 &&&&&& \\[-0.3cm]
Events at $25 \, \mathrm{fb}^{-1}$	                    & $0.2$ & $3.8$ &$0.4$ &$ $ & $$ & $$ & $$ & $$ \\[0.02cm]
\hline \hline
\end{tabular}
} \\[0.1cm]
\caption{Cross sections in~fb for the $4$ $b$-tag case at $8$~TeV. In the event line backgrounds are summed. For $t\bar{t}b\bar{b}$, the contribution of $t\bar{t}h$ is shown in parentheses.}
\label{tab:4b backgrounds8}
\end{center}
\end{table}

\begin{table}[t!]
\begin{center}
{\small
\begin{tabular}{c|cc|c|ccccc}
\hline \hline
\vspace{-0.35cm}\\
		 & \multicolumn{2}{c}{Signal} &\multicolumn{5}{|c}{Backgrounds} \\[0.02cm]
		
 \hline
  &&&&&&& \\[-0.3cm]
	Cuts	& $c_{F}=1$ & $c_{F}= -1$ & Total & $t Z \bar{b} j$ & $ t b \bar{b} \bar{b} j$ & $t \bar{t} b \bar{b}$  &  $t \bar{t} b \bar{b}$ (\emph{mis}) &  $t \bar{t} j$  \\[0.02cm]
		
 \hline
 &&&&&& \\[-0.3cm]
 Acceptance Cuts + $\epsilon$        			    & $0.19$ & $2.85$ &$39.14$ &$0.46$ & $1.07 $ & $14.40\,(1.94)$ & $11.53$ & $11.69 $\\[0.02cm]

$|m_{b b} - m_{h}| < 15$ GeV  			& $0.17$ & $2.61$ &$19.78$ &$0.12 $ & $0.32 $ & $4.88\,(1.63)$ & $6.52$ & $7.93$ \\[0.02cm]

min $m_{b b} > 90$ GeV  		    		& $0.13$ & $1.82$ &$5.97$ &$0.05$ & $0.09$ & $1.68\,(1.04)$ & $3.54$ & $0.61 $ \\[0.02cm]

min $m_{bj} > 170$ GeV 		                    		& $0.07$ & $1.20$ &$0.35$ &$0.02$ & $0.06 $ & $0.03\,(0.03)$ & $0.05$ & $0.19$ \\[0.02cm]
\hline
 &&&&&& \\[-0.3cm]
Events at $25 \, \mathrm{fb}^{-1}$	                    & $1.7$ & $30.1$ & $8.7$ &$$ & $ $ & $$ & $$ & $$  \\[0.03cm]
\hline \hline
\end{tabular}
} \\[0.1cm]
\caption{Cross sections in~fb for the $4$ $b$-tag case at $14$~TeV. In the event line backgrounds are summed. For $t\bar{t}b\bar{b}$, the contribution of $t\bar{t}h$ is shown in parentheses.}
\label{tab:4b backgrounds14}
\end{center}
\end{table}

The exclusion limits obtained for $c_{F}=-1$, assuming $25\,\mathrm{fb}^{-1}$ of data, are $2.4 \,\sigma$ and $\sim 6 \, \sigma$ at 8 and 14~TeV respectively. The sensitivity at 8~TeV is comparable to the one obtained in the $3$ $b$ case, while at 14~TeV requiring an extra $b$-jet improves the result significantly.

Before discussing the implications of our results, we wish to comment here on the sensitivity of the proposed analysis to the $t\bar{t}h$ process. As can be read from Tables~\ref{tab:4b backgrounds8} and \ref{tab:4b backgrounds14}, this process makes up a sizable fraction of the $t\bar{t}b\bar{b}$ cross section after the cuts. Moreover, after the first three cuts, the rate for $t\bar{t}h$ is comparable to the $th$ signal for $c_{F}=-1$. Being insensitive to the sign of the top Yukawa, $t\bar{t}h$ can be considered as a background process in our analysis. It is, however, quite useful to observe that the simple search strategy we propose in the $4b$ channel would be sensitive to both single and pair top production in association with a Higgs boson. In this respect, a key role is played by the cut on $m_{bj}$ that was designed to suppress processes with a $t\bar{t}$ pair in the final state, as discussed above. The relative contribution of $t\bar{t}h$ to the $t\bar{t}b\bar{b}$ background with one mistag, on the other hand, is small, approximately $5\%$.


\section{Implications on Higgs couplings}\label{sec:implications}

We are now able to study the implications of our results on the general parameter space of Higgs couplings. To do so we combine the two analyses that we discussed in Section~\ref{sec:signals}, i.e. $3$ and $4$ $b$-tags, to exploit the full LHC sensitivity in $th\to t b\bar{b}$ production. Note that in the combination we consider the $3b$ and $4b$ samples as independent. While this is an approximation (which can be easily lifted in a more realistic analysis by defining fully exclusive samples), in practice it has a small effect as the $4b$ sample is significantly smaller than the $3b$ one. We combine the (Poisson) $p$-values through Fisher's method, defining
\begin{equation}
  X^2=-2\sum_{i=1}^k \log{p_i}
\end{equation}
where $k=2$ in our case, and $p_{1,2}$ are the $p$-values of the two analyses. It can be shown that $X^2$ has a $\chi^{2}$ distribution with $2k$ degrees of freedom. Thus the combined $p$-value is the one associated to the value of $X^2$ at each point in parameter space. This definition is conservative compared to estimate based on the naive product of $p$-values.

\begin{figure}[t]
 \begin{center}
\includegraphics[width=.4\textwidth]{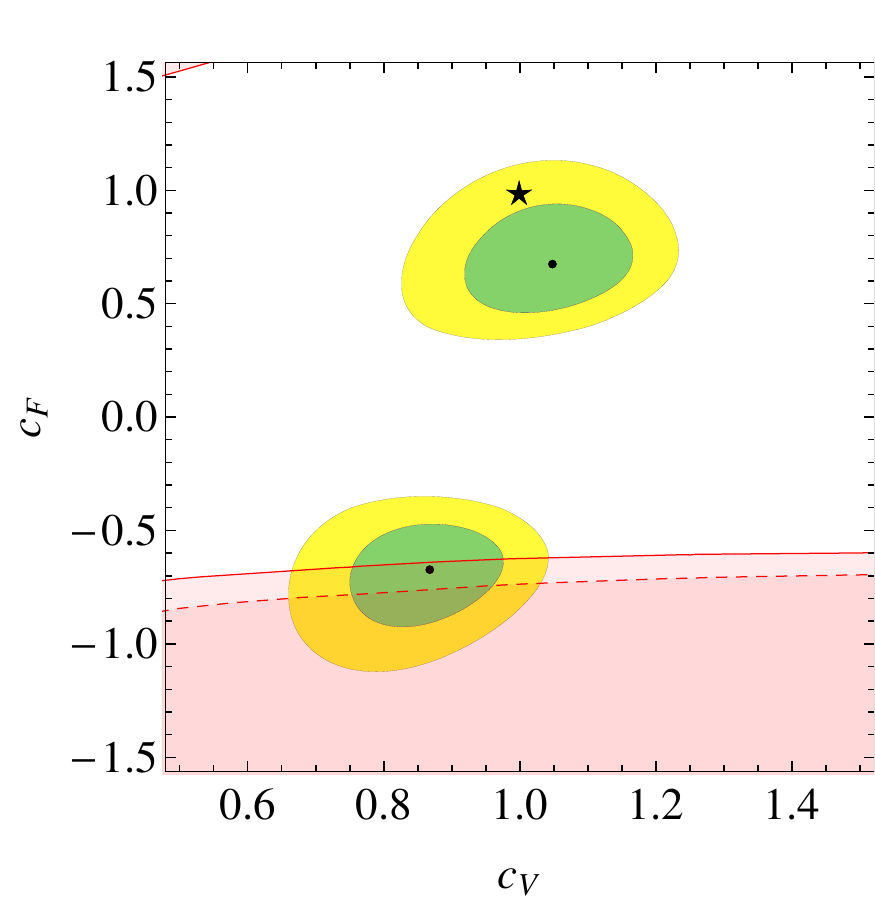}\hspace{1cm} \includegraphics[width=.4\textwidth]{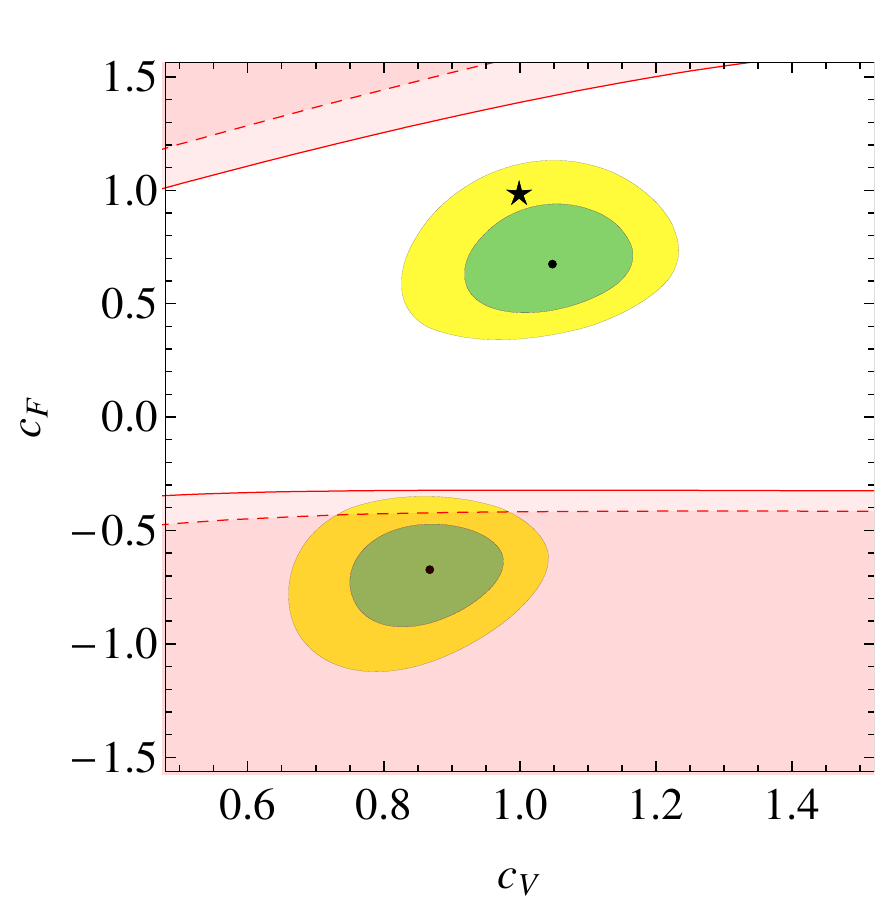}
\vspace{-.5cm}
 \end{center}
 \caption{Regions of the ($c_{V},c_{F}$) plane excluded at $95\%$ CL by our analysis of $th\to hb\bar{b}$ ($3$ and $4$ $b$ final states combined), at $8$~TeV (left) and $14\,\mathrm{TeV}$ (right), assuming an integrated luminosity of $25\,\mathrm{fb}^{-1}$ and $50\,\mathrm{fb}^{-1}$ (dashed and solid respectively). The $68\%$ and $95\%$ CL contours of a fit to current Higgs data are also shown, in green and yellow respectively. A universal rescaling by $c_{F}$ of the Higgs coupling to fermions is assumed. The Higgs coupling fit is based on the data reported by ATLAS, CMS and Tevatron after ICHEP 2012 and collected in Ref.~\cite{Espinosa:2012im}.}
\label{fig:bestfit}
\end{figure}

In Fig.~\ref{fig:bestfit} we present the results of our analysis in the $(c_V,\,c_F)$ plane, where a universal rescaling of the Higgs couplings to fermions $c_{t}=c_{b}=c_{\tau}=c_{c}=c_{F}$ is assumed. The regions that can be excluded (at $95\%$ CL) by $th$ production with an integrated luminosity of $25$ and $50\,\mathrm{fb}^{-1}$ are presented, along with the regions currently favored by a fit to Higgs data. As can be seen, already at $8$~TeV parts of the preferred region with $c_{F}<0$ can be excluded. The current best fit point with $c_{F}<0$ is excluded at $2.1\,\sigma$ with $50 \, \mathrm{fb}^{-1}$. On the other hand, a moderate luminosity at $14$~TeV can conclusively remove the degeneracy between the two regions that are at the present time preferred by Higgs data, for example reaching a $5.8 \,\sigma$ exclusion of the best fit point with $c_{F}<0$ after $50 \, \mathrm{fb}^{-1}$. Notice that in addition to the $th$ production cross section (recall Fig.~\ref{fig:sigmaRatio}), also the branching ratio of the Higgs into $b\bar{b}$ depends on the parameters $(c_{V},c_{F})$.

\begin{figure}[t]
 \begin{center}
\includegraphics[width=.4\textwidth]{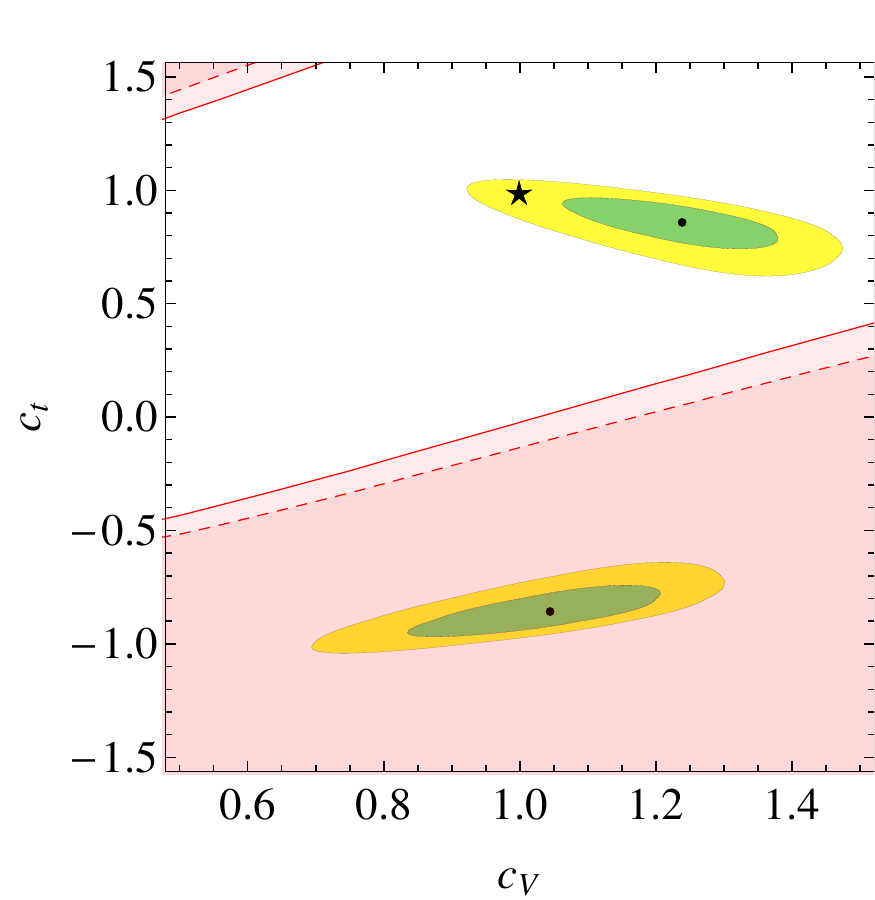}\hspace{1cm}\includegraphics[width=.4\textwidth]{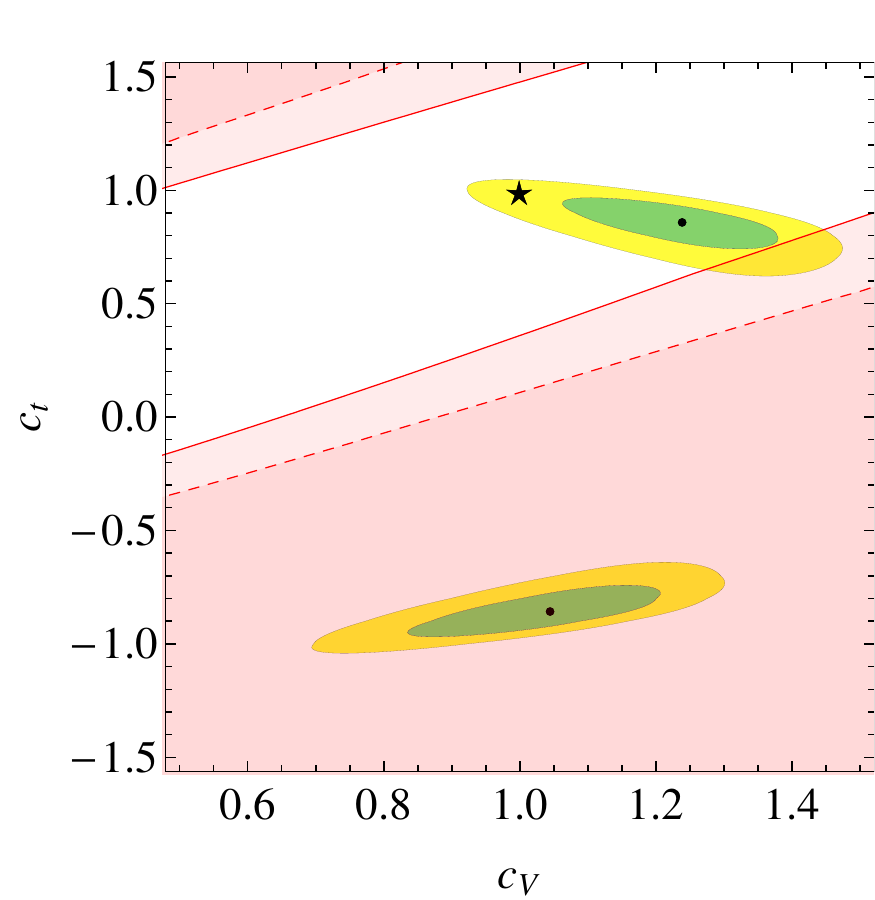}
 \vspace{-.5cm}
 \end{center}
 \caption{Regions of the ($c_{V},c_{t}$) plane excluded at $95\%$ CL by our analysis of $th\to hb\bar{b}$ ($3$ and $4$ $b$ final states combined), at $8$~TeV (left) and $14\,\mathrm{TeV}$ (right), assuming an integrated luminosity of $25\,\mathrm{fb}^{-1}$ and $50\,\mathrm{fb}^{-1}$ (dashed and solid respectively). The $68\%$ and $95\%$ CL contours of a fit to current Higgs data are also shown, in green and yellow respectively. The top Yukawa is assumed to be rescaled by $c_{t}$, while we have set $c_{b}=c_{\tau}=c_{c}=1$. The Higgs coupling fit is based on the data reported by ATLAS, CMS and Tevatron after ICHEP 2012 and collected in Ref.~\cite{Espinosa:2012im}.}
\label{fig:fitFixedBR}
\end{figure}

It is also possible to relax the assumption of universal couplings of the Higgs to fermions and consider the case where only the $ht\bar{t}$ coupling $c_t$ has a rescaled value compared to the SM while $c_{b}=c_{\tau}=c_{c}=1$, so in particular $\Gamma(h \to b \bar{b})$ is equal to its SM value. In this case, the $th\to tb\bar{b}$ rate is essentially fixed by the dependence on $c_{V},c_{t}$ of the production cross section (a mild sensitivity to $c_{V},c_{t}$ through the Higgs total width is also present). The results are shown in the $(c_{V}, c_{t})$ plane in Fig.~\ref{fig:fitFixedBR}. Excluded regions at 95\% C.L. are displayed for $25$~fb$^{-1}$ and $50$~fb$^{-1}$ integrated luminosity. Superimposed are the regions currently favored by Higgs data. The most striking feature is, that the best fit region with $c_{t} < 0$ can already be completely excluded at $8$~TeV with $25$~fb$^{-1}$ (reaching a $4.0\,\sigma$ exclusion of the best fit point with negative $c_{t}$).

\section{Conclusions}\label{sec:conclusions}

After the time of discovery comes the need for measuring. The couplings of the putative Higgs boson are of prime importance since they control the behavior of the whole theory at high energy. The dominant processes involving the Higgs boson that are currently investigated at the LHC do not allow us to determine all its couplings unambiguously. An important task now is therefore to systematically identify additional processes that could complement the first LHC information and lift degeneracies appearing in Higgs coupling fits.

In this paper, we have studied single top production in association with a Higgs boson, focusing on the Higgs decay into $b \bar{b}$. We discussed the form of the amplitude of the hard scattering process $W b \rightarrow t h$, showing that for nonstandard couplings of the Higgs to the $W$ boson and/or to the top quark a striking enhancement of the cross section can be obtained. The enhancement is due to the non-cancellation of terms that grow with energy in the amplitude and lead to violation of perturbative unitarity at some UV scale. We estimate the cutoff scale to be at least $10\,\mathrm{TeV}$, concluding that corrections to our computation of the cross section from physics above the cutoff are always negligible.

We have performed a parton-level study of the LHC signal processes $pp \rightarrow thj$ and $pp \rightarrow thjb$, and of the corresponding irreducible and some of the most relevant reducible backgrounds.  The combination of the two final states, containing $3$ and $4$ $b$-jets respectively, shows that if a universal rescaling $c_{F}$ of the fermion couplings is assumed, already at $8$~TeV parts of the preferred region with $c_{F}<0$ can be excluded. On the other hand, a moderate luminosity of $50 \, \mathrm{fb}^{-1}$ at $14$~TeV can conclusively remove the degeneracy between the two regions that are at the present time preferred by Higgs data, reaching a $5.8 \,\sigma$ exclusion of the best fit point with negative $c_{F}$. In addition, we investigated the case where only the $h t \bar{t}$ coupling differs from its SM value while the other Yukawa couplings are standard. Here, the best fit region with negative top Yukawa coupling can be completely excluded at $8$~TeV with $25$~fb$^{-1}$, reaching a $4.0\,\sigma$ exclusion of the best fit point with $c_{t}<0$.

Our results therefore motivate the undertaking of a full-fledged analysis by the ATLAS and CMS collaborations on one side, and the improvement on the accuracy of the theoretical predictions on the other. In the former case, in addition to having a complete simulation of $th$ events, one could also study the possibility of improving the signal over background ratio by using further discriminating variables (such as for example the different rates for $th$ and $\bar t h$ with respect to the main backgrounds which are symmetric) or multivariate analyses.
On the latter, it would be certainly interesting to evaluate the (possibly significant) impact of NLO QCD corrections to signal and irreducible backgrounds, i.e., $thj$ and $tZj$, a task that can now be accomplished in a fully automatic way~\cite{Hirschi:2011pa,Frederix:2009yq, Frederix:2011zi, Frederix:2012dh}.

Further information on the Higgs couplings to heavy quarks could also come from other processes at the LHC.
One  example is double Higgs production, $gg \to hh$. This process proceeds through a triangle and a box diagram, which, again, interfere destructively in the SM  and therefore result in a sensitive probe of the Higgs-heavy quarks interactions, see, e.g., Refs.~\cite{Grober:2010yv,Contino:2012xk,Gillioz:2012se}.
Finally we remark that complementary information could a priori also come from the observation of $B_s \to \mu^+ \mu^-$ very recently reported by LHC$b$~\cite{LHCb}. The measured value of BR$(B_s \to \mu^+ \mu^-)$ agrees well with the SM prediction~\cite{Buras:2012ru}. The SM contribution is actually dominated by the interactions associated to the top Yukawa coupling and therefore this measurement could be naively expected to provide a good probe of any deviation of the top Yukawa itself. However, only the Yukawa interactions between the Goldstone bosons and the quarks contribute to this process. What we have proposed to probe via $th$ production is rather the interaction of the physical Higgs boson with the top quark, i.e. the one controlled by the parameter $c_t$. Actually, if the deviations from $c_t=1$ originate from pure Higgs non-linearities as in composite Higgs models, for instance via a higher dimensional operator like $|H|^2 \bar{Q}_L H^\dagger t_R$, then it is easy to see that the prediction for BR$(B_s \to \mu^+ \mu^-)$ remains unaffected.\footnote{We thank G.~Isidori for illuminating discussions on this point.}

\section*{Note added}
During the final stages of this project another study discussing $th$ production appeared~\cite{Biswas:2012bd} that focuses on the $h\to \gamma\gamma$ decay channel.
\section*{Acknowledgments}
We are grateful to A.~Coccaro, P.~Francavilla, G.~Isidori and V.~Sanz for useful discussions. We also wish to thank V.~Hirschi for help with {\sc aMC@NLO}. M.F. thanks CERN for hospitality during the early stages of this work. This research has been partly supported by the European Commission under the ERC Advanced Grant 226371 {\it MassTeV} and the contract PITN-GA-2009-237920 {\it UNILHC}. C.G. is also supported by the Spanish Ministry MICNN under contract FPA2010-17747. The work of F.M. is partially supported by the IAP Program, BELSPO
VII/37 and the IISN-FNRS conventions 4.4511.10 and 4.4517.08. E.S. has been supported in part by the European Commission under the ERC Advanced Grant 267985 {\it DaMeSyFla}. The work of  M.F. has been supported in part by the NSF grant PHY-0757868 and by the \emph{Fondazione A.~Della Riccia}. A.T. has been partially supported by the Swiss National Science Foundation under contract 200020-138131.

\appendix

\section{Forward $Wb\to th$ scattering} \label{App: fw scatt}
%
The forward cross section for the partonic process $Wb\to th$, defined for example by a cut on $|\eta|>\bar{\eta}$, can be computed for large $s$ in a very simple way. In fact, for this purpose the diagram with top exchange in the $s$-channel can be neglected, and we only need to look at the diagram with $W$ exchange in the $t$-channel. In the regime we are interested in, i.e. large $s$, the longitudinal polarization of the $W$ dominates. The leading term in the amplitude, which is enhanced at small $|t|$, goes as $\sim s/(t-m_{W}^{2})\,$ and reads
\begin{equation} \label{fw amp}
\mathcal{A}^{fw}_{L} \simeq \frac{g\, c_{V}\, m_{W}}{\sqrt{2}v}\frac{1}{t-m_{W}^{2}}\bar{u}(p_{t})\slashed{p_{W}}(1-\gamma_{5})u(p_{b})
\end{equation}
At large $s$ and generic $t$, the fermion bilinears relevant to the amplitude read
\begin{align}
\bar{u}(p_{t})(1-\gamma_{5})u(p_{b})& = 2\sqrt{s}\, A\left(t/s,\varphi;\xi_{t},\xi_{b}\right) +2m_{t} B\left(t/s,\varphi;\xi_{t},\xi_{b}\right) + \ldots \\
\bar{u}(p_{t})\slashed{p_{W}}(1-\gamma_{5})u(p_{b})& = 2s\, B\left(t/s,\varphi;\xi_{t},\xi_{b}\right)+ \ldots
\end{align}
where the functions $A,B$ have been defined in Eqs.~(\ref{A}-\ref{B}), and the dots stand for subleading terms. Thus squaring the amplitude in Eq.~\eqref{fw amp}, summing and averaging over polarizations (we neglect the contributions of the transverse components of the $W$) and recalling that we are interested in the region $s \gg |t|$ we find
\begin{equation} \label{fw amp squared}
\overline{\left|\mathcal{A}^{fw}\right|^{2}}= \frac{g^{2}c_{V}^{2}m_{W}^{2}}{3v^{2}}\left(\frac{s}{t-m_{W}^{2}}\right)^{2}
\end{equation}
from which we derive the approximate expression of the forward cross section
\begin{equation} \label{forward cs}
\sigma(|\eta|>\bar{\eta}, s) \simeq \frac{c_{V}^{2}g^{2}}{48\pi v^{2}} R(\bar{\eta},s)\,,\qquad R(\bar{\eta},s) = \frac{(s/2m_{W}^{2})(1- \tanh \bar{\eta})}{1+(s/2m_{W}^{2})(1-\tanh \bar{\eta})}\,
\end{equation}
valid for $\tanh \bar{\eta} \approx 1$ (i.e. for large $\bar{\eta}$). 
We note that as expected, the forward cross section is controlled only by $c_{V}$ and is insensitive to the value of $c_{F}$. As a consequence, the forward cross section is insensitive to the growth with energy of the ``hard scattering'' amplitude, which takes place for $c_{V}\neq c_{F}$ and was discussed in Sec.~\ref{sec:main}. As a numerical example, let us consider $c_{V}=1$, a cut $|\eta|>3$ and let us set the center of mass energy to $\sqrt{s} = 5\,\mathrm{TeV}$. Then computing the cross section without approximations gives
\begin{equation}
\sigma_{\mathrm{full}}(|\eta|>3) = \{16.3,\,16.5,\,16.8\}\,\mathrm{pb}, \qquad \mathrm{for}\qquad c_{F} = \{1,\,0,\,-1\}
\end{equation}
whereas using the approximate formula in \eqref{forward cs} yields $\sigma(|\eta|>3, \sqrt{s}=5\,\mathrm{TeV}) = 16.4\,\mathrm{pb}$, a very accurate result. The factor $R$ has the value $R(\bar{\eta}=3,\, \sqrt{s} = 5\,\mathrm{TeV}) \simeq 0.91$.

%
%
%
%

\end{document}